\documentclass[12pt]{article}\usepackage{QMR-QC}
\begin{document}

\title{Use of a Quantum Computer\\
and the Quick Medical Reference\\
To Give an Approximate Diagnosis}

\author{Robert R. Tucci\\
        P.O. Box 226\\
        Bedford,  MA   01730\\
        tucci@ar-tiste.com}

\date{ \today}

\maketitle

\vskip2cm
\section*{Abstract}
The Quick Medical Reference (QMR) is
a compendium of statistical knowledge
connecting diseases to findings (symptoms).
The information in QMR
can be represented as a Bayesian network.
The inference problem
(or, in more medical language,
giving a diagnosis)
for the QMR is
to, given some findings,
find the probability of each
disease.
Rejection sampling and likelihood weighted
sampling (a.k.a. likelihood weighting) are two simple
algorithms for making approximate inferences
from an arbitrary Bayesian net (and from the QMR Bayesian net
in particular). Heretofore, the samples for these
two algorithms have been obtained with a conventional
``classical computer". In this paper,
we will show that two analogous algorithms
exist for the QMR Bayesian net,
where the samples are obtained
with a quantum computer.
We expect that these two algorithms,
implemented on a quantum computer,
can also be used to make inferences
(and predictions) with other
Bayesian nets.

\section{Introduction}

Trying to make inferences based on incomplete,
uncertain knowledge is a common
everyday problem. Computer
scientists have found that this problem can be
handled admirably well using Bayesian networks
(a.k.a. causal probabilistic networks)\cite{Jordan}.
Bayesian nets allow one to pose
and solve the inference
problem in a graphical fashion that
possesses a high degree of intuitiveness,
naturalness, consistency, reusability, modularity,
generality and simplicity.

This paper was motivated
by a series of papers written by me,
in which I define some nets
 that describe quantum phenomena.
I call them
``quantum Bayesian nets"(QB nets).
They are a counterpart to the
conventional ``classical Bayesian
nets" (CB nets) that describe
classical phenomena.
In particular, this
paper gives an example
of a general technique,
first proposed in Ref.\cite{Tuc00},
of embedding CB nets within QB nets.
The reader can
understand this paper easily
without having to read Ref.\cite{Tuc00}
first.
He might consult
Ref.\cite{Tuc00}
if he wants to understand
better the motivation
behind the constructs used
in this paper and how
they can be generalized.

\begin{figure}[h]
    \begin{center}
    \epsfig{file=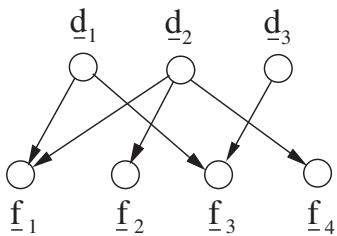, height=1.3in}
    \caption{Bayesian network of
    the same form as the QMR Bayesian network,
    but with considerably fewer parent (``diseases")
    and children (``findings") nodes.}
    \label{fig:cbnet-general}
    \end{center}
\end{figure}

The Quick Medical Reference (QMR) is
a compendium of statistical knowledge
connecting diseases to symptoms.
The original version of QMR
was compiled by Miller et al\cite{QMR-heur}.
Shwe et al\cite{QMR-prob}
designed  a CB net
based on the information of Ref.\cite{QMR-heur}.
The QMR CB net of Shwe et al
is of the form shown in Fig.\ref{fig:cbnet-general}.
It contains two layers:
a top layer of $\approx$600 parent nodes corresponding
to distinct diseases, and
a bottom layer of $\approx$4,000 children nodes corresponding
to distinct findings.
The inference problem
(or, in more medical language,
giving a diagnosis)
for the QMR is
to, given some findings,
find the probability of each
disease, or at least the more
likely diseases.

Making an inference with a CB
net usually requires
summing over the states of
a subset $S$ of the set of nodes of the
graph. If each node in $S$ contains
just 2 states, a sum over all the
states of $S$ is a sum over $2^{|S|}$
terms. These sums of exponential size
are the bane of the Bayesian network
formalism. It has been
shown that making exact\cite{complex-exact}
(or even approximate\cite{complex-approx})
inferences with a general CB net is
NP-hard.
In 1988,
Lauritzen and Spiegelhalter(LS)
devised a technique\cite{LauSpi} for
making inferences with
CB nets for
which the subset $S$
is relatively small
(for them, $S=S_{LS} =$ the maximal
clique of the moralized graph).
This led to a resurgence in
the use of CB nets,
as it allowed the use of nets
that hitherto had been prohibitively
expensive computationally.
According to Ref.\cite{Jaa99},
for the QMR CB net, $|S_{LS}| \approx 150$,
so the LS technique does not
help in this case.
Researchers have
found(Ref.\cite{Jaa99}
gives a nice review of their work)
many exact and approximate algorithms
for making inferences from the QMR CB net.
Still,
all currently known algorithms
require performing an
exponential number of operations.

Rejection sampling and likelihood weighted
sampling (a.k.a. likelihood weighting) are two simple
algorithms for making approximate inferences
from an arbitrary CB net (and from the QMR CB net
 in particular). Heretofore, the samples for these
 two algorithms have been obtained with a conventional
 ``classical computer". In this paper,
 we will show that two analogous algorithms
 exist for the QMR CB net,
 where the samples are obtained
 with a quantum computer.
 We will show that obtaining
 each sample, for these
 two algorithms, for the QMR CB net,
 on a quantum computer, requires only
 a polynomial number of steps.
 We expect that these two algorithms,
implemented on a quantum computer,
 can also be used to make inferences
 (and predictions) with other
CB nets.

\section{Notation}

In this section, we will
define some notation that is
used throughout this paper.
For additional information about our
notation, we recommend that
the reader
consult Ref.\cite{Paulinesia}.
Ref.\cite{Paulinesia} is
a review article, written
by the author of this paper, which
uses the same notation as this paper.

Let $Bool =\{0, 1\}$. As usual, let $\ZZ, \RR, \CC$ represent the set
of integers (negative  and non-negative),
real numbers, and
complex numbers, respectively.
For integers $a$, $b$
 such that $a\leq b$, let
$Z_{a,b}=\{a, a+1,
\ldots b-1, b\}$.
For any set $S$, let $|S|$
be the number of elements in $S$.
The power set of $S$, i.e.,
the
set of all subsets of $S$ (including
the empty and full sets),
will be denoted by $2^S$.
Note that $|2^S| = 2^{|S|}$.

We will use $\theta(S)$
to represent the ``truth function";
$\theta(S)$ equals 1 if statement $S$ is true
and 0 if $S$ is false.
For example, the Kronecker delta
function is defined by
$\delta^y_x=\delta(x,y) = \theta(x=y)$.

Random variables will be represented
by underlined letters.
For any random variable $\rvx$,
$val(\rvx)$ will denote the set
of values that $\rvx$ can assume.
Samples of $\rvx$ will be denoted
by $\sam{x}{k}$ for $k\in Z_{1, N_{sam}}$.

Consider an n-tuple $\vecf=(f_1,f_2, \ldots, f_n)$,
and a set $A\subset Z_{1,n}$.
By $(\vecf)_A$ we will mean $(f_i)_{i\in A}$
;
that is,
the $|A|$-tuple
that one creates from $\vecf$,
by keeping only the components
listed in $A$. If $\vecf\in Bool^n$,
then we will use the statement
$\vecf = 0$ to indicate that all
components of $\vecf$ are 0.
Likewise, $\vecf = 1$
will mean all its components are 1.

For any matrix $A\in\CC^{p\times q}$,
$A^*$ will stand for its complex
conjugate, $A^T$ for its transpose, and
$A^\dagger$ for its Hermitian conjugate.
When we write a matrix, and
leave some of its entries
blank, those blank entries
should be interpreted as zeros.

For any set $\Omega$ and any
function $f:\Omega\rarrow \RR$,
we will use
$f(x)/(\sum_{x\in\Omega} numerator)$
to mean
$f(x)/(\sum_{x\in\Omega} f(x))$.
This notation is convenient when
$f(x)$ is a long expression
that we do not wish to write twice.

Next we explain our notation for
quantum circuit diagrams.
We label single qubits (or qubit
positions) by a Greek letter or by an
integer. When we use integers,
the topmost qubit wire is 0, the next one
down is 1, then
2, etc.
{\it Note that in our
quantum circuit diagrams,
time flows from the right to the left
of the diagram.} Careful:
Many workers in Quantum
Computing draw their diagrams
so that time flows in the
opposite direction. We eschew their
convention because
it forces one to reverse
the order of the operators
every time one wishes to convert
between a circuit
diagram
and its algebraic equivalent
in Dirac notation.

\section{The QMR CB Net}

In this section, we
describe the QMR CB net.

\begin{figure}[h]
    \begin{center}
    \epsfig{file=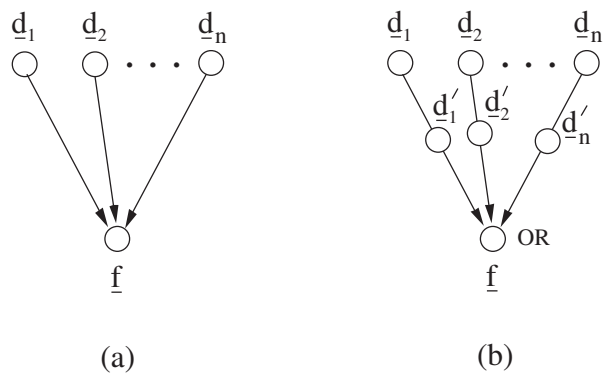, height=2in}
    \caption{
    $(a)$ CB net
with $n$ parent nodes (``diseases")
for
all
pointing into a single child node (``finding").
$(b)$
Noisy-or CB net, a special case or
an approximation of the CB net of Figure (a).
    }
    \label{fig:cbnet-nd-1f}
    \end{center}
\end{figure}

Before describing the QMR CB net,
let us describe the  noisy-or CB net
(invented by Pearl in Ref.\cite{Pearl}).
Consider a CB net
of the form of Fig.\ref{fig:cbnet-nd-1f}(a),
consisting of $n$ parent nodes (``diseases"),
$\rvd_j\in Bool$ with $j\in Z_{1,n}$,
all
pointing into a single child node (``finding"),
$\rvf\in Bool$.
The CB net of Fig.\ref{fig:cbnet-nd-1f}(a)
represents a probability distribution
$P(f,\vec{d})$ that satisfies:

\beq
P(f,\vec{d}) = P(f| \vecd)\prod_{j=1}^n P(d_j)
\;.
\label{eq:noisy-or-1}
\eeq
We
say the probability distribution of
Eq.(\ref{eq:noisy-or-1})
and Fig.\ref{fig:cbnet-nd-1f}(a)
is a {\bf noisy-or} if it also satisfies:

\begin{subequations}
\label{eq:noisy-or-2}
\beq
P(f,\vec{d}) =\left\{\sum_{\vec{d'}}
P(f|\vec{d'}) \prod_j
P(d'_j|d_j)
\right\} \prod_jP(d_j)
\;,
\eeq
with

\beq
P(f|\vec{d'}) = \delta(f, d'_1\vee d'_2\vee\dots \vee d'_n)
\;.
\eeq
\end{subequations}
For example, when $n=2$,

\beqa
P(f|\vec{d'})
&=&\left\{
\begin{array}{r}
d'_1,d'_2 \rarrow\\
\begin{array}{llllll}
&&\vline 00 &01&10&11 \\
\hline
f\downarrow
&0&\vline 1&0&0&0\\
&1&\vline 0&1&1&1\\
\end{array}
\end{array}
\right.
\\
&=& \delta(f, d'_1\vee d'_2)
\;.
\label{eq:or-of-2}
\eeqa
Eqs.(\ref{eq:noisy-or-2})
are represented by Fig.\ref{fig:cbnet-nd-1f}(b).
Sometimes,
one also restricts
 the distributions
$P(d'_j|d_j)$ to have the special
form:

\beqa
P(d'_j|d_j)&=&\left\{
\begin{array}{r}
d_j \rarrow\\
\begin{array}{llll}
&&\vline 0 &1 \\
\hline
d'_j\downarrow
&0&\vline 1& 1-q_{1j}\\
&1&\vline 0& q_{1j}\\
\end{array}
\end{array}
\right.
\\
&=&
(1-q_{1j})^{d_j}\delta_{d'_j}^0+
(q_{1j}d_j)\delta_{d'_j}^1
\;,
\label{eq:pd'd}
\eeqa
where $q_{1j}\in[0,1]$.
A general distribution
$P(d'_j|d_j)$ would
contain 2 degrees of
freedom whereas Eq.(\ref{eq:pd'd})
contains only one, namely $q_{1j}$.
Note that

\beq
P(f=0|\vecd) = \prod_j(1-q_{1j})^{d_j}
= e^{-\sum_j\theta_{1j}d_j}
\;,
\eeq
where $\theta_{1j} = -\ln(1-q_{1j})\geq 0$.
The inference problem for the noisy-or CB net
consists in calculating
$P(\vecd | f=0)$ and
$P(\vecd | f=1)$; that is, the probability
of diseases having the value $\vecd$,
given that $f$ is 0 or 1.
This is given by Bayes rule:

\beqa
P(\vecd | f=0) &=&
\frac{
P(f=0|\vecd)\prod_j P(d_j)
}{
P(f=0)
}
\\
&=&
\frac{\prod_j
\left\{
(1-q_{1j})^{d_j}P(d_j)
\right\}
}{
\sum_{\vecd} \;numerator
}
\label{eq:d-sum-for-noisy-or-net}
\;.
\eeqa
Note that the sum in the denominator of
Eq.(\ref{eq:d-sum-for-noisy-or-net})
is over $2^n$ terms.

Now that we understand the noisy-or CB net,
it's easy to understand the QMR CB net.
The QMR CB net
consists of multiple noisy-or CB nets,
one for each finding.
Suppose the QMR CB net
has $N_D$ diseases (parent nodes),
$\rvd_j\in Bool$ for $j\in Z_{1, N_D}$,
and
$N_F$ findings (children nodes),
$\rvf_i\in Bool$ for $i\in Z_{1, N_F}$.
Then,  for each $i\in Z_{1, N_F}$,
one has\footnote{It's possible to include
``leakage" in the definitions of
noisy-or and QMR nets, but we
won't include it
since it can be ignored without
loss of generality.
One can add a leakage node $\rvL_i\in Bool$
pointing into each $\rvf_i$ node,
for each $i\in Z_{1,N_F}$.
These leakage nodes behave
just like disease nodes
that are always ``turned on"(i.e., set to
1).
Then,
instead of Eq.(\ref{eq-fi-given-pa}),
one has

\beq
P(f_i=0| L_i=1,(\vecd)_{pa(\rvf_i)})=
(1-q_{i0})
\prod_{j\in pa(\rvf_i)}
\left\{
(1-q_{ij})^{d_j}
\right\}
=
e^{-\theta_{i0}-\sum_{j\in pa(\rvf_i)}\theta_{ij}d_j}
\;.
\nonumber
\eeq
}

\beq
P(f_i=0|(\vecd)_{pa(\rvf_i)})=
\prod_{j\in pa(\rvf_i)}
\left\{
(1-q_{ij})^{d_j}
\right\}
=
e^{-\sum_{j\in pa(\rvf_i)}\theta_{ij}d_j}
\;,
\label{eq-fi-given-pa}
\eeq
where $q_{ij}\in [0,1]$
and $pa(\rvf_i)\subset Z_{1,N_D}$
is the set of parents of node $\rvf_i$.
Let $I_0$, $I_1$ and $I_{unk}$ constitute
a disjoint partition of $Z_{1,N_F}$.
(``unk" stands for unknown.)
The inference problem for the QMR CB net
consists in calculating
$P[\vecd|(\vecf)_{I_0}=0, (\vecf)_{I_1}=1] $.
By Bayes rule,

\begin{eqnarray}
P[\vecd|(\vecf)_{I_0}=0, (\vecf)_{I_1}=1] &=&
\frac{P[(\vecf)_{I_0}=0, (\vecf)_{I_1}=1|\vecd]P(\vecd)}
{P[(\vecf)_{I_0}=0, (\vecf)_{I_1}=1]}\\
&=&
\frac{
\Pi_1\Pi_0P(\vecd)
}{
P_{I_1,I_0}
}
\label{eq:d-sum-for-QMR-net}
\;,
\end{eqnarray}
where

\beq
\Pi_0 =
\prod_{i\in I_0}\prod_{j\in pa(\rvf_i)}
\left\{
(1-q_{ij})^{d_j}
\right\}
\;,
\label{eq:def-Pi-0}
\eeq

\beq
\Pi_1 =
\prod_{i\in I_1}
\left\{
1 - \prod_{j\in pa(\rvf_i)}
\left\{
(1-q_{ij})^{d_j}
\right\}
\right\}
\;,
\label{eq:def-Pi-1}
\eeq
and

\beqa
P_{I_1,I_0}&=&
P[(\vecf)_{I_0}=0, (\vecf)_{I_1}=1]\\
&=&
\sum_\vecd
P[(\vecf)_{I_0}=0, (\vecf)_{I_1}=1,\vecd]
\;.
\label{eq:p-d-sum}
\eeqa

Note that the numerator
of Eq.(\ref{eq:d-sum-for-QMR-net})
can be calculated in a polynomial number
of steps,
but its denominator
(i.e.,  $P_{I_1,I_0}$)
is expressed in Eq.(\ref{eq:p-d-sum})
as a sum over $2^{N_D}$ terms.
Calculating
$P_{I_1,I_0}$ naively, by summing
numerically those $2^{N_D}$ terms, is
unfeasible when $N_D$ is large.

At the end of this paper are
4 appendices.
Reading them is not
a prerequisite to
understanding the rest of
this paper, but
they might be of
interest to some readers.

In Appendix \ref{app:d-sum},
we show that

\beq
P_{I_1,I_0}=
\sum_{S\subset I_1}
(-1)^{|S|}
T_{S,I_0}
\;,
\label{eq-t-to-p-mobius}
\eeq
where
$T: 2^{I_1}\times 2^{I_0}\rarrow \RR$
is some function that can be calculated
in a polynomial number of steps.
Thus, $P_{I_1,I_0}$ can be calculated by
summing numerically over $2^{|I_1|}$ terms,
regardless of $|I_0|$ size.
This is better than
$2^{N_D}$ terms, but still unfeasible
for $|I_1|$
large.

Eq.(\ref{eq-t-to-p-mobius})
 can be inverted. For more
on this see Appendix \ref{app-mobius}.

Rejection sampling and likelihood weighted
sampling are two
algorithms for making approximate inferences
from an arbitrary CB net (and from the QMR CB net
 in particular).
 Heretofore, the samples for these
 two algorithms have been obtained with a conventional
 ``classical computer".
 In case the reader is not
 familiar with these two algorithms,
 in the manner they
 have  been implemented heretofore on a classical computer,
 see Appendices \ref{app-reject-sam}
 and \ref{app-like-sam}
 for an introduction to them.
  In the next section,
 we will show that two analogous algorithms
 exist for the QMR CB net,
 where the samples are obtained
 with a quantum computer.

\section{Diagnosis Via Quantum Computer}
In this section,
we will describe a method
for making inferences from
the QMR using a quantum computer.

A slight change of notation:
the parameter $q_{ij}\in [0,1]$
of the previous section will
be replaced in this section
by a sine squared. Let

\beq
q_{ij} = \sin^2{\alpha_{ij}} = S^2_{ij}\;\;,\;\;
1-q_{ij} = \cos^2{\alpha_{ij}} = C^2_{ij}
\;,
\eeq
for some real number $\alpha_{ij}$.
We will also abbreviate $S_{1j}$
by $S_j$ and $C_{1j}$
by $C_j$.

\begin{figure}[h]
    \begin{center}
    \epsfig{file=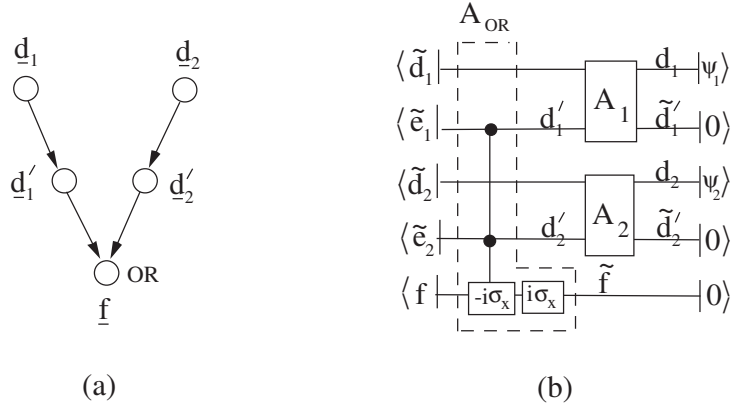, height=2.2in}
    \caption{
$(a)$ CB net
consisting of two diseases
pointing to one finding.
$(b)$
Quantum circuit that
generates
some of the same probability
distributions as the
CB net of Figure(a).}
    \label{fig:qbnet-2d-1f}
    \end{center}
\end{figure}

We begin by considering the
simple case of a CB net
consisting of two diseases
pointing to one finding,
as displayed in Fig.\ref{fig:qbnet-2d-1f}(a).
We will next show that
Fig.\ref{fig:qbnet-2d-1f}(b)
is a quantum circuit that
can generate
some of the same probability
distributions as the
CB net Fig.\ref{fig:qbnet-2d-1f}(a).
The state vectors
$\ket{\psi_1}, \ket{\psi_2}$, and the
unitary transformations
$A_1,A_2, A_{OR}$
that appear in
the quantum circuit
of Fig.\ref{fig:qbnet-2d-1f}(b)
are defined as follows.

For $j=1,2$, define
$\ket{\psi_j}$ by

\beq
\ket{\psi_j}=
U_j \ket{0}
\;,
\eeq
where

\beq
U_j =
\left[
\begin{array}{cc}
\sqrt{P_{\rvd_j}(0)} & -\sqrt{P_{\rvd_j}(1)}\\
\sqrt{P_{\rvd_j}(1)} & \sqrt{P_{\rvd_j}(0)}
\end{array}
\right]
\;,
\;\;
\ket{0}
=
\left[
\begin{array}{c}
1\\
0
\end{array}
\right]
\;.
\eeq

For $j=1,2$, let

\beqa
A_j(d'_j, \td_j|\tpd_j, d_j)
&=&\left\{
\begin{array}{r}
\tpd_j,d_j \rarrow\\
\begin{array}{llllll}
&&\vline 00 &01&10&11 \\
\hline
&00&\vline 1& 0&0&0\\
d'_j,\td_j\downarrow
&01&\vline 0& C_j&0&-S_j\\
&10&\vline 0& 0&1&0\\
&11&\vline 0& S_j&0&C_j\\
\end{array}
\end{array}
\right.
\label{eq:2-bit-q-embed}
\\
&=&
\left[
(C_j^{d_j})\delta^{d_j}_{\td_j}
\delta^0_{d'_j}
+
(S_jd_j)\delta^{d_j}_{\td_j}
\delta^1_{d'_j}
\right]
\delta^0_{\tpd_j}
+
\left[
\dots
\right]
\delta^1_{\tpd_j}
\;.
\eeqa
For those familiar with
Ref.\cite{Tuc00}, note that the probability amplitude
$A_j(d'_j, \td_j|\tpd_j, d_j)$
is a q-embedding of the probability
distribution
$P(d'_j|d_j)$
defined in Eq.(\ref{eq:pd'd}).
Note also that source and sink nodes are
denoted by letters with tildes over them.

The matrix given by
Eq.(\ref{eq:2-bit-q-embed}) is a 2 qubit
unitary transformation. Such
transformations can be
decomposed (compiled)
into an expression containing at most
3 CNOTs, using a method due
to Vidal and Dawson\cite{VD}
(For software that performs this
decomposition, see Ref.\cite{Tuc05}).

Let

\beqa
A_{OR}(f, \te_1, \te_2| \tf, d'_1, d'_2)
&=&\left\{
\begin{array}{r}
\tf,d'_1,d'_2 \rarrow\\
\begin{array}{llllllllll}
&&\vline 000 &001&010&011&100 &101&110&111 \\
\hline
&000&\vline 1&&&&0&&&\\
f,\te_1,\te_2\downarrow
&001&\vline&0&&&&i&&\\
&010&\vline&&0&&&&i&\\
&011&\vline&&&0&&&&i\\
&100&\vline 0&&&&1&&&\\
&101&\vline&i&&&&0&&\\
&110&\vline&&i&&&&0&\\
&111&\vline&&&i&&&&0\\
\end{array}
\end{array}
\right.
\label{eq:quan-or-matrix}
\\
&=&
[i^f\delta_f^{d'_1\vee d'_2}
\delta_{d'_1,d'_2}^{\te_1,\te_2}]
\delta_\tf^0
+[\dots]\delta_\tf^1
\;.
\eeqa
For those familiar with
Ref.\cite{Tuc00}, note that the probability amplitude
$A_{OR}(f, \te_1, \te_2| \tf, d'_1, d'_2)$
is a q-embedding of the probability
distribution
$P(f| d'_1, d'_2)$
defined in Eq.(\ref{eq:or-of-2}).

The matrix given by
Eq.(\ref{eq:quan-or-matrix}) can be compiled
as follows:

\beqa
[A_{OR}(f, \te_1, \te_2| \tf, d'_1, d'_2)]&=&
e^{i\frac{\pi}{2}\sigx\otimes
\sum_{(b,b')\in Bool^2-(0,0)}P_{b,b'}}\\
&=&
e^{i\frac{\pi}{2}\sigx\otimes I_4}
e^{-i\frac{\pi}{2}\sigx\otimes P_{00}}\\
&=&
i\sigx(2)[-i\sigx(2)]^{n(0)n(1)}
\;.
\eeqa

The probability
$P(f, \te_1,\te_2, \td_1,\td_2)$
for the quantum circuit
Fig.\ref{fig:qbnet-2d-1f}(b)
is given by:

\beqa
\lefteqn{P(f, \te_1,\te_2, \td_1,\td_2)=}
\nonumber
\\
&=&
\left|
\sum_{d'_1,d'_2,d_1,d_2}
A_{OR}(f,\te_1,\te_2|\tf=0,d'_1,d'_2)
\prod_{j=1,2}
\left\{A_j(d'_j,\td_j|\tpd_j=0,d_j)
\sqrt{P(d_j)}\right\}
\right|^2
\\
&=&
\left|
\sum_{d'_1,d'_2,d_1,d_2}
i^f
\delta_f^{d'_1\vee d'_2}
\delta_{d'_1,d'_2}^{\te_1,\te_2}
\prod_{j=1,2}
\left\{[(C_j^{d_j})
\delta_{\td_j}^{d_j}
\delta_{d'_j}^0
+
(S_jd_j)
\delta_{\td_j}^{d_j}
\delta_{d'_j}^1]
\sqrt{P(d_j)}\right\}
\right|^2
\;.
\eeqa
In particular, when $f=0$,

\beq
P(f=0, \te_1,\te_2, \td_1,\td_2)=
\prod_{j=1,2}
C_j^{2\td_j}
P(\td_j)
\delta_{\te_j}^0
\;.
\eeq
If
$\te_1$ and $\te_2$
are not observed, we may sum over them to get

\beq
P(f=0, \td_1,\td_2)=
\prod_{j=1,2}
C_j^{2\td_j}
P(\td_j)
\;.
\eeq
If we replace $\td_j$ by $d_j$,
$P(f=0, \td_1,\td_2)$
for
the quantum circuit
Fig.\ref{fig:qbnet-2d-1f}(b)
is identical
to
$P(f=0, d_1,d_2)$
for
the CB net
Fig.\ref{fig:qbnet-2d-1f}(a).
This is no coincidence.
The quantum
circuit was designed
from the CB net
to make this true.
In a sense defined
in Ref.\cite{Tuc00},
the CB net is embedded
in the quantum circuit.

\begin{figure}[h]
    \begin{center}
    \epsfig{file=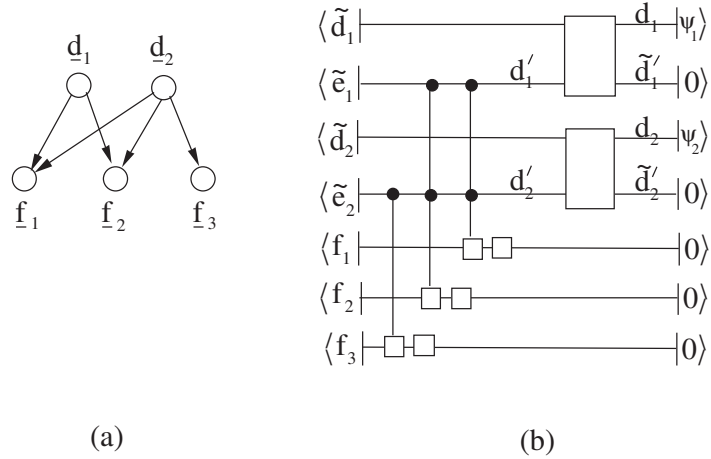, height=2.5in}
    \caption{
$(a)$ QMR-like CB net
with two diseases
and three finding.
$(b)$
Quantum circuit that
generates
some of the same probability
distributions as the
CB net of Figure(a).}
    \label{fig:qbnet-2d-3f}
    \end{center}
\end{figure}

One can easily generalize
this example with $N_D=2$
and $N_F=1$ to arbitrary
$N_D$ and $N_F$.
Fig.\ref{fig:qbnet-2d-3f}
gives an example with
$N_D=2$
and $N_F=3$.
In the example with
$N_D=2$ ,$N_F=1$,
we set:

\beq
[A_{OR}(f, \te_1, \te_2| \tf, d'_1, d'_2)]=
i\sigx(2)[-i\sigx(2)]^{n(0)n(1)}
\;.
\eeq
For arbitrary $N_D,N_F$,
this equation can be
generalized to:

\beq
[A_{OR}(f_i,\{\te_j\}_{j\in pa(\rvf_i)}|
\tf_i,\{d'_j\}_{j\in pa(\rvf_i)}]=
i\sigx(\tau_i)[-i\sigx(\tau_i)]^{
\prod_{\kappa\in K_i}n(\kappa)}
\;,
\eeq
for $i\in Z_{1, N_F}$,
where $\tau_i$ is the
qubit label of
qubit $\rvf_i$,
and  $K_i$ is the
set of
qubit labels for
the parents of qubit $\rvf_i$.

For arbitrary $N_D,N_F$,
we can generalize this construction to
obtain  a quantum circuit
that yields probabilities
$P(\vecf,\vec{\te},\vec{\td})$.
If the external outputs
$\vec{\te}$ are not observed,
then we measure
$P(\vecf,\vec{\td})$.
If we replace $\vec{\td}$ by $\vecd$,
the probability
$P(\vecf,\vec{\td})$
for the quantum circuit
is identical to the
probability
$P(\vecf,\vecd)$
for the CB net that was
embedded in that quantum
circuit. As
discussed previously, the inference problem
for the CB net is to find
$P[\vecd|(\vecf)_{I_0}=0, (\vecf)_{I_1}=1]$.
This probability equals
$P[(\vecf)_{I_0}=0, (\vecf)_{I_1}=1,\vecd]$
divided by
$P[(\vecf)_{I_0}=0, (\vecf)_{I_1}=1]$.
The numerator
$P[(\vecf)_{I_0}=0, (\vecf)_{I_1}=1,\vecd]$
can be calculated exactly numerically
on a conventional classical computer.
Not so the denominator
$P[(\vecf)_{I_0}=0, (\vecf)_{I_1}=1]$,
at least not for large $|I_1|$.
Here is where the quantum computer
shows its mettle.
One can run the quantum circuit
many times, in either of two modes,
to get a so called empirical distribution
that approximates
$P[\vecd | (\vecf)_{I_0}=0, (\vecf)_{I_1}=0]$.
The empirical distribution
converges to the exact
one. The two modes that
we are referring to are
rejection sampling
and likelihood weighted sampling.
We describe each of these separately
in the next two sections.

\subsection{Rejection Sampling}

Assume that we are given
the number of samples $N_{sam}$
that we intend to collect,
and the sets $I_0,I_1,I_{unk}$
which are a disjoint partition of
$Z_{1,N_F}$.
Then the rejection sampling
algorithm goes as
follows (expressed in pseudo-code,
pidgin C language):

\fbox{\parbox{5in}{
\begin{verse}
For all $\vecd$ $\{W(\vecd)=0;\}$\\
$W_{tot}=0;$\\
For samples $k=1,2, \ldots,N_{sam}\{$\\
\hspace{2em}Generate $(\sam{\vecd}{k},\sam{\vecf}{k})$ with quantum computer;\\
\hspace{2em}If $(\sam{\vecf}{k})_{I_0}=0$ and $(\sam{\vecf}{k})_{I_1}=1 \{$(//rejection here\\
\hspace{4em}If $\sam{\vecd}{k} = \vecd \{ W(\vecd)++;\}$ \\
\hspace{4em}$W_{tot}++;$\\
\hspace{2em}$\}$\\
$\}// k$ loop (samples)\\
For all $\vecd$ $\{P[\vecd|(\vecf)_{I_0}=0,(\vecf)_{I_1}=1 ]=\frac{W(\vecd)}{W_{tot}};\}$\\
\end{verse}
}}

A convergence proof of this
algorithm goes as follows.
For any function
$g:Bool^{N_D + N_F}\rarrow \RR$,
as
$N_{sam}\rarrow \infty$,
the sample average
$\overline{g(\sam{\vecd}{k},\sam{\vecf}{k})}$
 tends  to:

\beq
\overline{g(\sam{\vecd}{k},\sam{\vecf}{k})} =
\frac{1}{N_{sam}}\sum_k g(\sam{\vecd}{k},\sam{\vecf}{k})
\rarrow \sum_{\vecd',\vecf'}P(\vecd',\vecf')g(\vecd',\vecf')
\;.
\eeq
Therefore,

\beqa
\frac{W(\vecd)}{W_{tot}}&=&
\frac{\frac{1}{N_{sam}}\sum_k \delta_{(\sam{\vecf}{k})_{I_0}}^0\delta_{(\sam{\vecf}{k})_{I_1}}^1\delta_{\sam{\vecd}{k}}^{\vecd}}
{\frac{1}{N_{sam}}\sum_k  \delta_{(\sam{\vecf}{k})_{I_0}}^0\delta_{(\sam{\vecf}{k})_{I_1}}^1}\\
&\rarrow&
\frac{\sum_{\vecd',\vecf'}P(\vecd',\vecf') \delta_{(\vecf')_{I_0}}^0\delta_{(\vecf')_{I_1}}^1\delta_{\vecd'}^{\vecd}}
{\sum_{\vecd',\vecf'}P(\vecd',\vecf') \delta_{(\vecf')_{I_0}}^0\delta_{(\vecf')_{I_1}}^1}\\
&\rarrow&
P[\vecd|(\vecf)_{I_0}=0,(\vecf)_{I_1}=1]
\;.
\eeqa

\subsection{Likelihood Weighted Sampling}

For likelihood weighted sampling,
the quantum circuit must be modified
as follows:
We assume that all
gates in the quantum circuit are elementary; that is, either single-qubit
transformations or controlled elementary
gates (like CNOTs or multiply-controlled
NOTs or multiply-controlled phases).

\begin{enumerate}
\item
For any qubit $\rvf_i$ with $i\in I_1$,
initialize the qubit to state $\ket{1}$.
(For any qubit $\rvf_i$ with $i\in I_0$,
initialize the qubit to state $\ket{0}$, same as
before.)
\item
For any qubit $\rvf_i$ with $i\in I_0\cup I_1$,
remove those elementary gates that can change
the state of $\rvf_i$. In particular,
remove any single-qubit gates acting on
$\rvf_i$,  and any controlled elementary gates that
use $\rvf_i$ as a target.
Do not remove controlled elementary gates that
use $\rvf_i$ as a control only.
\end{enumerate}

Assume that we are given
the number of samples $N_{sam}$
that we intend to collect,
and the sets $I_0,I_1,I_{unk}$
which are a disjoint partition of
$Z_{1,N_F}$.
Then the likelihood weighted sampling
algorithm goes as
follows (expressed in pseudo-code,
pidgin C language):

\fbox{\parbox{6in}{
\begin{verse}
For all $\vecd$ $\{W(\vecd)=0;\}$\\
$W_{tot}=0;$\\
For samples $k=1,2, \ldots,N_{sam}\{$\\
\hspace{2em}{\footnotesize Generate $(\sam{\vecd}{k},\sam{\vecf}{k})$  subject to $(\vecf)_{I_0}=0,(\vecf)_{I_1}=1$ with quantum computer;}\\
\hspace{2em}$L= \prod_{i\in I_0}P[\sam{f_i}{k}=0|(\sam{\vecd}{k})_{pa(\rvf_i)}]
\prod_{i\in I_1}P[\sam{f_i}{k}=1|(\sam{\vecd}{k})_{pa(\rvf_i)}];$\\
\hspace{2em}If $\sam{\vecd}{k} = \vecd \{ W(\vecd)\;+=\;L;\}$ \\
\hspace{2em}$W_{tot}\;+=\;L;$\\
$\}// k$ loop (samples)\\
For all $\vecd$ $\{P[\vecd|(\vecf)_{I_0}=0,(\vecf)_{I_1}=1 ]=\frac{W(\vecd)}{W_{tot}};\}$\\
\end{verse}
}}

A convergence proof of this
algorithm goes as follows.
Define the likelihood functions
$L_{evi}$ and $L_{unk}$
by (``evi" stands for evidence
and ``unk" for unknown):

\beq
L_{evi}(\vecd) =
\prod_{i\in I_0}P[f_i=0|(\vecd)_{pa(\rvf_i)}]
\prod_{i\in I_1}P[f_i=1|(\vecd)_{pa(\rvf_i)}]
\;,
\eeq
and

\beq
L_{unk}(\vecd,\vecf) =
\prod_{i\in I_{unk}}P[f_i|(\vecd)_{pa(\rvf_i)}]
\;.
\eeq
Clearly,

\beq
P(\vecd,\vecf) = L_{evi}(\vecd)L_{unk}(\vecd,\vecf)P(\vecd)
\;.
\eeq
For any function
$g:Bool^{N_D + N_F}\rarrow \RR$, as
$N_{sam}\rarrow \infty$,
the sample average
$\overline{g(\sam{\vecd}{k},\sam{\vecf}{k})}$
 tends to:

\beq
\overline{g(\sam{\vecd}{k},\sam{\vecf}{k})} =
\frac{1}{N_{sam}}\sum_k g(\sam{\vecd}{k},\sam{\vecf}{k})
\rarrow \sum_{\vecd',\vecf'}
\delta_{(\vecf')_{I_0}}^0\delta_{(\vecf')_{I_1}}^1
L_{unk}(\vecd',\vecf')P(\vecd')
g(\vecd',\vecf')
\;.
\eeq
Therefore,

\beqa
\frac{W(\vecd)}{W_{tot}}&=&
\frac{\frac{1}{N_{sam}}\sum_k L_{evi}(\sam{\vecd}{k})\delta_{\sam{d}{k}}^{\vecd}}
{\frac{1}{N_{sam}}\sum_k L_{evi}(\sam{\vecd}{k})}\\
&\rarrow&
\frac{\sum_{\vecd',\vecf'}P(\vecd',\vecf') \delta_{(\vecf')_{I_0}}^0\delta_{(\vecf')_{I_1}}^1\delta_{\vecd'}^{\vecd}}
{\sum_{\vecd',\vecf'}P(\vecd',\vecf') \delta_{(\vecf')_{I_0}}^0\delta_{(\vecf')_{I_1}}^1}\\
&\rarrow&
P[\vecd|(\vecf)_{I_0}=0,(\vecf)_{I_1}=1]
\;.
\eeqa

\begin{appendix}

\section{Appendix:
Summing $P[(\vecf)_{I_0}=0, (\vecf)_{I_1}=1,\vecd]$\\
over $\vecd$} \label{app:d-sum}

In this appendix, we will
sum $P[(\vecf)_{I_0}=0, (\vecf)_{I_1}=1,\vecd]$\\
over $\vecd$. This is like performing
a multidimensional integral.

Recall that $P_{I_1,I_0}$ was defined as:

\beq
P_{I_1,I_0}=
\sum_\vecd
P[(\vecf)_{I_0}=0, (\vecf)_{I_1}=1,\vecd]
\;.
\eeq
For all $i\in Z_{1,N_F}$ and
$j\in Z_{1,N_D}$, let

\beq
(\vech_i)_j =
\;\;\theta_{ij}\theta[j\in pa(\rvf_i)]\;\;=
\left\{
\begin{array}{l}
 \theta_{ij}\;\;{\rm if} \;\;j\in pa(\rvf_i)\\
 0\;\;{\rm otherwise}
 \end{array}
 \right.
\;.
\eeq
For all $j\in Z_{1,N_D}$,
we can always find $\alpha_j, \beta_j\in \RR$
so that $P(d_j)$ can be expressed as:

\beq
P(d_j) = e^{-\alpha_j - \beta_j d_j}
\;.
\eeq
Now $\Pi_0$ (defined by Eq.(\ref{eq:def-Pi-0})),
$\Pi_1$ (defined by Eq.(\ref{eq:def-Pi-1})),
and $P(\vecd)$
can be expressed as:

\beq
\Pi_0 =
\prod_{i\in I_0}
e^{- \vech_i\cdot\vecd}
\;,
\eeq

\beq
\Pi_1 =
\prod_{i'\in I_1}\left\{
1-e^{- \vech_{i'}\cdot\vecd}
\right\}
\;,
\eeq
and

\beq
P(\vecd)=
e^{-\alpha - \vec{\beta}\cdot \vecd}
\;,{\rm where}\;\;
\alpha = \sum_{j=1}^{N_D} \alpha_j
\;.
\eeq
Thus

\beqa
P_{I_1,I_0}&=&
\sum_\vecd
\Pi_0\Pi_1 P(\vecd)\\
&=&
e^{-\alpha}
\sum_\vecd
e^{-\vec{\beta}\cdot \vecd -
\sum_{i\in I_0}\vech_i\cdot\vecd}
\prod_{i'\in I_1}
\left\{
1 - e^{-\vech_{i'}\cdot \vecd}
\right\}
\;.
\label{eq:p-before-prod-to-sum}
\eeqa
Consider any set $\Omega$
and any function
$f:\Omega \rarrow \RR$.
When $\Omega = \{a,b\}$,

\beq
(1- e^{-f(a)})
(1- e^{-f(b)})=
1 -  e^{-f(a)} -  e^{-f(b)}
+  e^{-f(a)-f(b)}
\;.
\eeq
This generalizes to

\beq
\prod_{x\in \Omega}\left\{
1 - e^{-f(x)}
\right\}=
\sum_{S\in 2^\Omega}
(-1)^{|S|}e^{-\sum_{x\in S}f(x)}
\;.
\label{eq:prod-to-sum}
\eeq
Using identity Eq.(\ref{eq:prod-to-sum}),
Eq.(\ref{eq:p-before-prod-to-sum})
yields

\beq
P_{I_1,I_0}=
e^{-\alpha}
\sum_{S\subset I_1}
(-1)^{|S|}
\sum_{\vecd}
e^{-\vec{\beta}\cdot\vecd
-\sum_{i\in I_0\cup S}\vech_i\cdot\vecd}
\;.
\label{eq:p-before-t-phi}
\eeq
For any $j \in Z_{1,N_D}$
and $A\subset Z_{1,N_F}$, define

\beq
\phi_j(A) =
\beta_j + \sum_{i\in A}(\vech_i)_j
\;.
\eeq
Also define a function $t:\RR\rarrow \RR$ by

\beq
t(\phi)=
\frac{1}{2}\sum_{d=0}^1 e^{-\phi d}
=\frac{1}{2}(1 + e^{-\phi})
\;.
\eeq
Using these definitions,
 Eq.(\ref{eq:p-before-t-phi})
yields

\beqa
P_{I_1,I_0}&=&
e^{-\alpha}
\sum_{S\subset I_1}
(-1)^{|S|}2^{N_D}
\prod_{j\in Z_{1,N_D}}
t[\phi_j(I_0\cup S)]
\\
&=&
\sum_{S\subset I_1}
(-1)^{|S|}
T_{S,I_0}
\;.
\eeqa
$T$ is a function
$T: 2^{I_1}\times 2^{I_0}
\rarrow \RR$
defined by the last equation.

\section{Appendix: Mobius Inversion Theorem
and Eq.(\ref{eq-t-to-p-mobius})}
\label{app-mobius}

In this Appendix, we discuss the
application of
 the Mobius Inversion Theorem\cite{Mobius} to
Eq.(\ref{eq-t-to-p-mobius}).

\begin{figure}[h]
    \begin{center}
    \epsfig{file=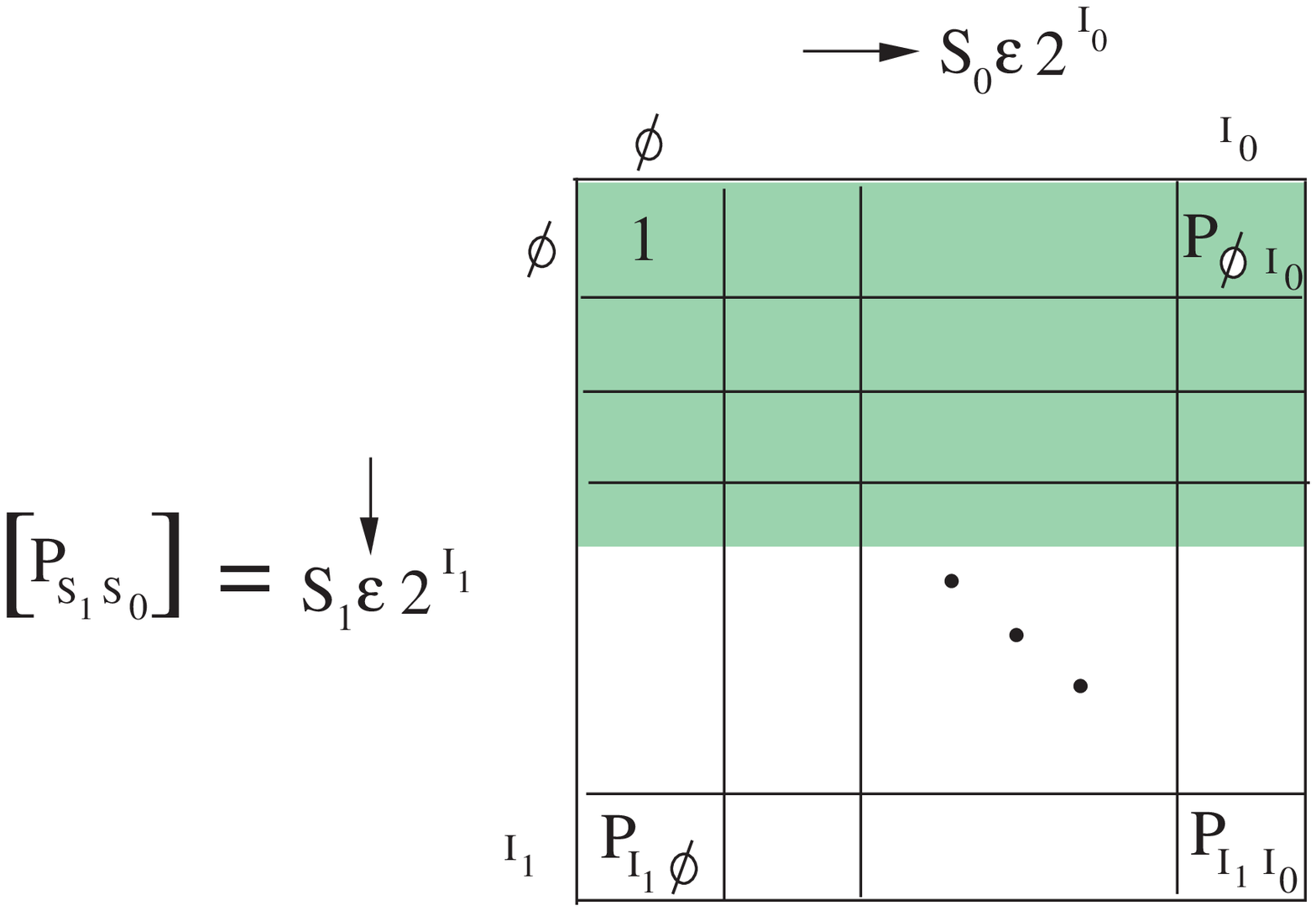, height=2.5in}
    \caption{
    The matrix $P_{S_1,S_0}=
P[(\vecf)_{S_1}=1, (\vecf)_{S_0}=0]$
for all $S_1\in 2^{I_1}$
and $S_0\in 2^{I_0}$.}
    \label{fig:p-s1-s0}
    \end{center}
\end{figure}

Fig.\ref{fig:p-s1-s0} shows
the matrix $P_{S_1,S_0}=
P[(\vecf)_{S_1}=1, (\vecf)_{S_0}=0]$
for all $S_1\in 2^{I_1}$
and $S_0\in 2^{I_0}$,
assuming large $|I_1|$ but
arbitrarily $|I_0|$.
We label the rows and
columns of $P_{S_1,S_0}$
in order of increasing set size.
The top-left corner entry is
$P_{\emptyset,\emptyset} =1$
and the bottom-right corner entry is
$P_{I_1,I_0}$.
Note that
$P_{S_1,S_0}\geq P_{I_1, I_0}$
for all $S_1\subset I_1$,
$S_0\subset I_0$.
The shaded
top part (corresponding to
 small or moderate $|S_1|$) of this matrix
can be calculated numerically
with a classical computer.
But not the unshaded bottom part
(corresponding to large $|S_1|$).
An empirical approximation
of the bottom part can be obtained
with a quantum computer.

Consider any set $J$ and any functions
$f, g : 2^J \rarrow \CC$.
The Mobius Inversion
Theorem\cite{Mobius} states that

\beq
g(J) =\sum_{J'\subset J}(-1)^{|J-J'|} f(J')
\;\;\;
\iff
\;\;\;
f(J) = \sum_{J'\subset J}
g(J')
\;.
\eeq
Using the fact that
when $J'\subset J$, $|J-J'| = |J|-|J'|$,
and replacing $g(J)$ by $(-1)^{|J|}g(J)$
in the previous equation, we get

\beq
g(J) =\sum_{J'\subset J}(-1)^{|J'|} f(J')
\;\;\;
\iff
\;\;\;
f(J) = \sum_{J'\subset J}
(-1)^{|J'|}
g(J')
\;.
\label{eq:mobius2}
\eeq
We
showed in Appendix \ref{app:d-sum}
that

\beq
P_{I_1,I_0} = \sum_{S_1\subset I_1}
(-1)^{|S_1|}T_{S_1,I_0}
\;.
\label{eq:T-alt-sum}
\eeq
Thus, by virtue of Eq.(\ref{eq:mobius2}),

\beq
T_{I_1,I_0} = \sum_{S_1\subset I_1}
(-1)^{|S_1|}P_{S_1,I_0}
\;.
\label{eq:P-alt-sum}
\eeq
More generally, if
$S'_1\subset I_1$,
$S_0\subset I_0$,
and

\beq
M_{S'_1, S_1} = (-1)^{S_1}
\;,
\eeq
then

\beq
P_{S'_1,S_0} = \sum_{S_1\subset S'_1}
M_{S'_1, S_1}T_{S_1,S_0}
\;,
\eeq
and

\beq
T_{S'_1,S_0} = \sum_{S_1\subset S'_1}
M_{S'_1, S_1}P_{S_1,S_0}
\;.
\eeq

Eq.(\ref{eq:P-alt-sum}) implies

\beq
P_{I_1,I_0} =
(-1)^{|I_1|}
\left\{
T_{I_1,I_0} -\sum_{S_1\subsetneqq I_1}
(-1)^{|S_1|}P_{S_1,I_0}
\right\}
\;.
\eeq
To approximate $P_{I_1,I_0}$,
one can estimate the right
hand side of the last equation.
$T_{I_1,I_0}$, and $P_{S_1,I_0}$
for small and moderate $|S_1|$,
can be calculated exactly
numerically on a classical
computer.
$P_{S_1,I_0}$
for large $|S_1|$
can be obtained empirically
on a quantum computer.

\section{Appendix:
Rejection Sampling \\for CB Nets
on a Classical Computer}\label{app-reject-sam}

In this Appendix, we review
the rejection sampling algorithm
for arbitrary CB nets on a classical computer.

Consider a CB net whose nodes
are labeled in topological order
by $\rvx_1, \rvx_2, \ldots \rvx_{N_{nds}}\equiv \rvx$.
Assume that $E$ (evidence set)
and $H$ (hypotheses set)
are disjoint subsets of $Z_{1,N_{nds}}$,
with $Z_{1,N_{nds}}-E\cup H$ not
necessarily empty.
Assume that we are given
the number of samples $N_{sam}$
that we intend to collect,
and the prior evidence $(x)_E$.
Then the rejection sampling
algorithm goes as
follows (expressed in pseudo-code,
pidgin C language):

\fbox{\parbox{5in}{
\begin{verse}
For all $(x)_H$ $\{W[(x)_H]=0;\}$\\
$W_{tot}=0;$\\
For samples $k=1,2, \ldots,N_{sam}\{$\\
\hspace{2em}For nodes $i=1,2, \ldots,N_{nds}\{$\\
\hspace{4em}Generate $\sam{x_i}{k}$ from $P[x_i|(\sam{x}{k})_{pa(\rvx_i)}];$\\
\hspace{4em}$//$Note that $pa(\rvx_i)\subset Z_{1, i-1}$ so \\
\hspace{4em}$//(\sam{x}{k})_{pa(\rvx_i)}$ has been calculated at this point\\
\hspace{2em}$\}// i$ loop (nodes)\\
\hspace{2em}If $(\sam{x}{k})_E = (x)_E \{$//rejection here\\
\hspace{4em}If $(\sam{x}{k})_H = (x)_H \{ W[(x)_H]++;\}$ \\
\hspace{4em}$W_{tot}++;$\\
\hspace{2em}$\}$\\
$\}// k$ loop (samples)\\
For all $(x)_H$ $\{P[(x)_H|(x)_E]=\frac{W[(x)_H]}{W_{tot}};\}$\\
\end{verse}
}}

A convergence proof of this
algorithm goes as follows.
For any function
$g:val(\rvx)\rarrow \RR$,
as
$N_{sam}\rarrow \infty$,
the sample average
$\overline{g(\sam{x}{k})}$ tends  to:

\beq
\overline{g(\sam{x}{k})} \equiv
\frac{1}{N_{sam}}\sum_k g(\sam{x}{k})
\rarrow \sum_{x'}P(x')g(x')
\;.
\eeq
Therefore,

\beqa
\frac{W[(x)_H]}{W_{tot}}&=&
\frac{\frac{1}{N_{sam}}\sum_k \delta[(x)_{E\cup H}, (\sam{x}{k})_{E\cup H}]}
{\frac{1}{N_{sam}}\sum_k \delta[(x)_E, (\sam{x}{k})_E]}\\
&\rarrow&
\frac{\sum_{x'}P(x') \delta[(x)_{E\cup H}, (x')_{E\cup H}]}
{\sum_{x'}P(x') \delta[(x)_E, (x')_E]}\\
&\rarrow&
\frac{P[(x)_{E\cup H}]}{P[(x)_E]}
\;.
\eeqa

\section{Appendix:
Likelihood Weighted Sampling \\for CB Nets
on a Classical Computer}
\label{app-like-sam}

In this Appendix, we review
the likelihood weighted sampling algorithm
for arbitrary CB nets
on a classical computer\cite{Fung, Shac}.

Consider a CB net whose nodes
are labeled in topological order
by $\rvx_1, \rvx_2, \ldots \rvx_{N_{nds}}\equiv \rvx$.
Assume that $E$ (evidence set)
and $H$ (hypotheses set)
are disjoint subsets of $Z_{1,N_{nds}}$,
with $Z_{1,N_{nds}}-E\cup H$ not
necessarily empty.
Let $X^c = Z_{1,N_{nds}}-X$ for any
$X\subset Z_{1,N_{nds}}$.
Assume that we are given
the number of samples $N_{sam}$
that we intend to collect,
and the prior evidence $(x)_E$.
Then the likelihood weighted sampling
algorithm goes as
follows (expressed in pseudo-code,
pidgin C language):

\fbox{\parbox{5in}{
\begin{verse}
For all $(x)_H$ $\{W[(x)_H]=0;\}$\\
$W_{tot}=0;$\\
For samples $k=1,2, \ldots,N_{sam}\{$\\
\hspace{2em}$L=1;$\\
\hspace{2em}For nodes $i=1,2, \ldots,N_{nds}\{$\\
\hspace{4em}If $i\in E^c\{$\\
\hspace{6em}Generate $\sam{x_i}{k}$ from $P[x_i|(\sam{x}{k})_{pa(\rvx_i)}];$\\
\hspace{6em}$//$Note that $pa(\rvx_i)\subset Z_{1, i-1}$ so \\
\hspace{6em}$//(\sam{x}{k})_{pa(\rvx_i)}$ has been calculated at this point\\
\hspace{4em}\}else if $i\in E$\{\\
\hspace{6em}$\sam{x_i}{k} = x_i;//(x)_E$ known\\
\hspace{6em}$L\;*=\;P[x_i|(\sam{x}{k})_{pa(\rvx_i)}];$\\
\hspace{4em}$\}$\\
\hspace{2em}$\}// i$ loop (nodes)\\
\hspace{2em}If $(\sam{x}{k})_H = (x)_H \{ W[(x)_H]\;+=\;L;\}$ \\
\hspace{2em}$W_{tot}\;+=\;L;$\\
$\}// k$ loop (samples)\\
For all $(x)_H$ $\{P[(x)_H|(x)_E]=\frac{W[(x)_H]}{W_{tot}};\}$\\
\end{verse}
}}

A convergence proof of this
algorithm goes as follows.
Define the likelihood function:

\beq
L_A(x) =
\prod_{i\in A}P[x_i|(x)_{pa(\rvx_i)}]
\;
\eeq
for any $A\subset Z_{1,N_{nds}}$.
Clearly,

\beq
P(x) = L_E(x)L_{E^c}(x)
\;.
\eeq
For any function
$g:val(\rvx)\rarrow \RR$, as
$N_{sam}\rarrow \infty$,
the sample average
$\overline{g(\sam{x}{k})}$ tends to:

\beq
\overline{g(\sam{x}{k})} =
\frac{1}{N_{sam}}\sum_k g(\sam{x}{k})
\rarrow \sum_{x'}
L_{E^c}(x')
\delta[(x)_E,(x')_E]
g(x')
\;.
\eeq
Therefore,

\beqa
\frac{W[(x)_H]}{W_{tot}}&=&
\frac{\frac{1}{N_{sam}}\sum_k L_E(\sam{x}{k})\delta[(x)_{H}, (\sam{x}{k})_{H}]}
{\frac{1}{N_{sam}}\sum_k L_E(\sam{x}{k})}\\
&\rarrow&
\frac{\sum_{x'}P(x') \delta[(x)_{E\cup H}, (x')_{E\cup H}]}
{\sum_{x'}P(x') \delta[(x)_E, (x')_E]}\\
&\rarrow&
\frac{P[(x)_{E\cup H}]}{P[(x)_E]}
\;.
\eeqa

\end{appendix}

\end{document}